\documentclass[aps,prl,twocolumn,superscriptaddress,showpacs]{revtex4}
\usepackage{graphicx}
\usepackage{latexsym}
\usepackage{amssymb}
\usepackage{amsmath}
\usepackage{amsfonts}
\usepackage{bm}
\usepackage{multirow}
\usepackage{color}

\newcommand{\vect}[1]{{\bm{#1}}}

\newcommand{\beq}{\begin{equation}}
\newcommand{\eeq}{\end{equation}}
\newcommand{\beqn}{\begin{eqnarray}}
\newcommand{\eeqn}{\end{eqnarray}}

\begin{document}

\title{Self-dual Quantum Electrodynamics as Boundary State of the three dimensional \\ Bosonic Topological Insulator}



\author{Cenke Xu}

\author{Yi-Zhuang You}

\affiliation{Department of physics, University of California,
Santa Barbara, CA 93106, USA}

\date{\today}

\begin{abstract}

Inspired by the recent developments of constructing novel Dirac
liquid boundary states of the $3d$ topological
insulator~\cite{maxashvin,wangsenthil1,wangsenthil2}, we propose
one possible $2d$ boundary state of the $3d$ bosonic symmetry
protected topological state with $U(1)_e \rtimes Z_2^T \times
U(1)_s $ symmetry. This boundary theory is described by a $(2+1)d$
quantum electrodynamics (QED$_3$) with two flavors of Dirac
fermions ($N_f = 2$) coupled with a noncompact U(1) gauge field: $
\mathcal{L} = \sum_{j = 1}^2 \bar{\psi}_j \gamma_\mu (\partial_\mu
- i a_\mu) \psi_j - i A^{s}_\mu \bar{\psi_i} \gamma_\mu
\tau^z_{ij} \psi_j + \frac{i}{2\pi} \epsilon_{\mu\nu\rho} a_\mu
\partial_\nu A^{e}_\rho $, where $a_\mu$ is the internal
noncompact U(1) gauge field, $A^s_\mu$ and $A^e_\mu$ are two
external gauge fields that couple to $U(1)_s$ and $U(1)_e$ global
symmetries respectively.
We demonstrate that this theory has a ``self-dual" structure,
which is a fermionic analogue of the self-duality of the
noncompact CP$^1$ theory with easy plane
anisotropy~\cite{ashvinlesik,deconfine1,deconfine2}. Under the
self-duality, the boundary action takes exactly the same form
except for an exchange between $A^s_\mu$ and $A^e_\mu$. The
self-duality may still hold after we break one of the U(1)
symmetries (which makes the system a bosonic topological
insulator), with some subtleties that will be discussed.

\end{abstract}


\maketitle

{\it --- 1. Introduction}

A symmetry protected topological (SPT) state may have very
different boundary states without changing the bulk state,
depending on the boundary Hamiltonian. As was shown in
Ref.~\onlinecite{TI_fidkowski1,TI_fidkowski2,TI_qi,TI_senthil,TI_max},
besides the well-known boundary state, $i.e.$ a single $2d$ Dirac
fermion, the boundary of an interacting $3d$ topological insulator
(TI) could have a topological order that respects all the
symmetries of the system but cannot be realized in a $2d$ system.
Very recently, the ``family" of boundary states of TI has been
even further expanded~\cite{maxashvin,wangsenthil1,wangsenthil2}:
it was shown that the boundary of the $3d$ TI could be a $(2+1)d$
quantum electrodynamics (QED$_3$) with one single flavor of
gauge-charged Dirac fermion, while the flux quantum of the U(1)
gauge field carries charge $1/2$ under the external
electromagnetic (EM) field $A_\mu$: \beqn \mathcal{L} = \bar{\psi}
\gamma_\mu (\partial_\mu - i a_\mu) \psi + \frac{1}{2}
\frac{i}{2\pi} \epsilon_{\mu\nu\rho} a_\mu
\partial_\nu A_\rho. \label{qed1}\eeqn This Lagrangian without
time-reversal symmetry can also be realized as a pure $2d$ theory
in the half-filled Landau level~\cite{son2015}. This boundary
state Eq.~\ref{qed1} is particularly interesting because it
suggests a duality between interacting $2d$ single Dirac fermion
and the noncompact QED$_3$ with one gauge-charged Dirac fermion,
which is further supported by the proof of S-duality in the $3d$
bulk~\cite{S}, and also the exact duality (or the mirror symmetry)
between certain $(2+1)d$ supersymmetric field
theories~\cite{mirror}. This duality is a very elegant analogue of
its standard bosonic version: the duality between the $(2+1)d$ XY
model and the bosonic QED$_3$ with one flavor of gauge-charged
complex boson~\cite{halperindual,leedual}. Although the infrared
fate of Eq.~\ref{qed1} under gauge fluctuation and fermion
interaction is unclear, it was demonstrated in
Ref.~\onlinecite{maxashvin,wangsenthil1,wangsenthil2} that
Eq.~\ref{qed1} can be viewed as the parent state of other
well-known boundary states of $3d$ TI.

In this work we will further extend the idea of
Ref.~\onlinecite{maxashvin,wangsenthil1,wangsenthil2}, and
construct novel boundary states of $3d$ bosonic SPT states. We
will consider bosons with a $U(1)_e \rtimes Z_2^T \times U(1)_s$
symmetry, where $U(1)_e$ can be viewed as the U(1) symmetry of the
electromagnetic charge, and $U(1)_s$ can be viewed as the spin
symmetry generated by total spin along $z$ direction. These
symmetries can be carried by a two component complex boson field
$z_\alpha$, which transforms under $U(1)_e$, $U(1)_s$ and
time-reversal as \beqn && U(1)_e: z_\alpha \rightarrow e^{i\theta}
z_\alpha, \ \ \ U(1)_s: z_\alpha \rightarrow \left( e^{- i \tau^z
\theta} \right)_{\alpha\beta} z_\beta, \cr\cr && \mathcal{T}:
z_\alpha \rightarrow (i\tau^y)_{\alpha\beta} z_\beta.
\label{z}\eeqn Notice that here the fact $\mathcal{T}^2 = -1$ can
be changed by a $U(1)_s$ rotation. It has been understood that for
a bosonic TI, the response to an eternal gauge field (either
$A^s_\mu$ or $A^e_\mu$ that couple to $U(1)_s$ and $U(1)_e$ global
symmetry) contains a $\theta \vect{E} \cdot \vect{B}/(4\pi^2)$
term with $\theta = \pm 2\pi$~\cite{senthilashvin}, which
corresponds to an integer quantum Hall state with $\sigma_{xy} =
\pm 1$ at the boundary. This is forbidden in a pure $2d$ bosonic
system without fractionalization~\cite{maxfisher}.

In this work we propose that the boundary of this bosonic SPT
state constructed with $z_\alpha$ can be a noncompact QED$_3$ with
two gauge-charged Dirac fermions: \beqn \mathcal{L} &=& \sum_{j =
1}^2 \bar{\psi}_j \gamma_\mu (\partial_\mu - i a_\mu) \psi_j
\cr\cr &-& i A^{s}_\mu \bar{\psi_i} \gamma_\mu \tau^z_{ij} \psi_j
+ \frac{i}{2\pi} \epsilon_{\mu\nu\rho} a_\mu
\partial_\nu A^{e}_\rho, \label{qed2}\eeqn where $\bar{\psi} = \psi^\dagger
\gamma^0$, and $\gamma^0 = \sigma^y$, $\gamma^1 = \sigma^x$,
$\gamma^2 = \sigma^z$. We will argue later that compared with
Eq.~\ref{qed1}, this theory with fermion flavor $N_f = 2$ has a
better chance to have a stable $(2+1)d$ conformal field theory
fixed point (perhaps with certain short range fermion
interaction), if we ignore the external gauge fields $A^s_\mu$ and
$A^e_\mu$. The coupling to the external gauge fields imply that
the spin symmetry $U(1)_s$ is carried by the fermionic fields
$\psi_j$, but the charge symmetry $U(1)_e$ is carried by the flux
of the noncompact gauge field $a_\mu$, which makes $a_\mu$ a
noncompact gauge field. We will also show that this theory has a
nice self-dual structure, the dual theory takes exactly the same
form as Eq.~\ref{qed2}, except for an exchange between the roles
of $A^e_\mu$ and $A^s_\mu$. This self-duality is reminiscent of
the more familiar self-duality of the noncompact CP$^1$ theory
with easy plane
anisotropy~\cite{ashvinlesik,deconfine1,deconfine2}, which
involves two gauge charged complex bosons and one noncompact gauge
field.

{\it --- 2. Microscopic construction of the noncompact QED$_3$
with $N_f = 2$}

In this section we will give a microscopic construction of
Eq.~\ref{qed2}. The starting point of our construction is similar
to Ref.~\onlinecite{maxashvin}: a $3d$ U(1) spin liquid state with
deconfined compact internal U(1) gauge field $a_\mu$ and fermionic
spinons $f_{j, \alpha}$ with gauge charge $+1$ that transforms as
\beqn \mathcal{T}: f_{j,\alpha} \rightarrow (i
\sigma^y)_{\alpha\beta} f^\dagger_{j,\beta}, \ \ \ U(1)_s: f_i
\rightarrow \left( \exp(i \theta \tau^z) \right)_{ij}f_j, \eeqn
under time-reversal and $U(1)_s$ global symmetry. $j = 1,2$ is a
flavor index that the symmetry $U(1)_s$ operates on.
$f_{j,\alpha}$ does not carry $U(1)_e$ charge. The U(1) gauge
symmetry and the time-reversal symmetry so-defined commute with
each other, thus this spin liquid has $U(1)_g \times Z_2^T$
``symmetry", where U(1)$_g$ stands for the U(1) gauge symmetry.
Now we put $f_{1,\alpha}$ and $f_{2,\alpha}$ both in a TI with
topological number $n = 1$. Notice that since here $f_\alpha$ has
$U(1)_g \times Z_2^T$ symmetry, at the mean field level the
classification of the spinon TI is
$\mathbb{Z}$~\cite{ludwigclass1,ludwigclass2,kitaevclass}, while
this classification can be reduced under
interaction~\cite{chenhe3B,senthilhe3}. The boundary of this spin
liquid is a QED$_3$ with two Dirac cones, but up until this point
the internal U(1) gauge field $a_\mu$ is propagating in the entire
$3d$ bulk.

Our next task is to confine the gauge field in the bulk, while
making the gauge field at the $2d$ boundary noncompact.
Ref.~\onlinecite{maxashvin} proposed a very nice way of achieving
this goal, which we will adopt here. Because $f_{1,\alpha}$ and
$f_{2,\alpha}$ each forms a $n = 1$ TI, a $2\pi-$monopole of
$a_\mu$ in the $3d$ bulk will acquire total polarization gauge
charge $+1$~\footnote{The magnetic monopole of $a_\mu$ is
invariant under time-reversal symmetry, while the polarization
charge is odd under time-reversal, thus whether this polarization
charge of the $2\pi-$monopole is $+1$ or $-1$ depends on how
time-reversal symmetry is broken on the boundary, which we have
confirmed numerically. But no matter how time-reversal symmetry is
broken at the boundary, we can always construct the Kramers
doublet boson $b_\alpha$ monopole we need with quantum numbers $(q
= 0, Q_s = \pm 1 , 2\pi)$.}, which comes from $+1/2$ polarization
density of $f_1$ and $f_2$ each. The quantum number of this
$2\pi-$monopole is $(q=1, Q_s = 0, 2\pi)$, where $q$ and $Q_s$
stand for the internal gauge charge and the $U(1)_s$ charge
respectively. Now by binding this monopole with a spinon $f$, we
obtain a gauge neutral object, and it is a boson which we call
$b$. Depending on whether we bind the monopole with $f_1$ or
$f_2$, $b$ can carry quantum number $(q = 0, Q_s = \pm 1, 2\pi)$,
thus $b$ is a doublet boson $b_\alpha$ with $\alpha = 1,2$. The
bosonic statistics of $b_\alpha$ comes from the fermionic
statistics of $f_\alpha$ and the mutual statistics between
$f_\alpha$ and the monopole.

There is another way of looking at the quantum number of the boson
doublet $b_\alpha$. A $2\pi$ monopole of $a_\mu$ could be viewed
as the source of a double-vortex of the superconductor of $f$,
since $f$ will view a single vortex as $\pi-$flux. Of course, we
need to consider a superconductor order parameter that preserves
the $U(1)_s$ symmetry. Then the source of a double vortex in this
system, will acquire four Majorana fermion zero modes, or
equivalently two complex fermion zero modes $f^0_{1}$ and $f^0_2$.
Our boson doublet states $b_1^\dagger |0\rangle$ (or $b_2^\dagger
|0\rangle$) corresponds to the states with filled (or unfilled)
$f^0_{1}$ zero mode and unfilled (or filled) $f^0_2$ zero mode.
Because each fermion zero mode will lead to $U(1)_s$ charge $\pm
1/2$ depending on whether it is filled or unfilled, $b_1^\dagger
|0\rangle$ and $b_2^\dagger |0\rangle$ will carry $U(1)_s$ charge
$\pm 1$ respectively. As was pointed out by
Ref.~\onlinecite{senthilhe3}, $b_\alpha$ is also a Kramers doublet
boson with $\mathcal{T}^2 = -1$. The fact $\mathcal{T}^2 = -1$ for
boson $b_\alpha$ can be derived by coupling these two zero modes
to a three component vector $\vect{N}$: $f^{0 \dagger} \vect{\tau}
f^0 \cdot \vect{N}$, after integrating out $f^0_j$, the effective
action for $\vect{N}$ is a $(0+1)d$ O(3) nonlinear sigma model
with a Wess-Zumino-Witten term at level-1~\cite{abanov2000}, whose
ground state is a Kramers doublet with $\mathcal{T}^2 = - 1$
because $\vect{N}$ is odd under time-reversal~\footnote{More
precisely, $\vect{N}$ is odd under the effective time-reversal
symmetry introduced in Ref.~\onlinecite{senthilhe3}, which is a
combination of $\mathcal{T}$ and a $\pi-$rotation of the pairing
order parameter.}.

Now let us take another Kramers doublet boson $z_\alpha$
introduced in Eq.~\ref{z} which carries both global $U(1)_e$ and
$U(1)_s$ charge , and form a time-reversal singlet bound state $D$
with $b_\alpha$: $D = (z_1 b_2 - z_2 b_1)$. $D$ carries total
quantum number $(Q_e = 1, q = 0, Q_s = 0, 2\pi)$. After condensing
this bound state $D$ in the bulk, $\mathcal{T}$ is still
preserved, while the $3d$ bulk is driven into a gauge confined
phase, because overall speaking $D$ carries a $2\pi$ monopole of
the internal gauge field, but it carries zero gauge charge, thus
all the spinons in the bulk are confined. Because the bound state
$D$ carries both the $U(1)_e$ charge and the $U(1)_g$ magnetic
monopole, its condensate does not break the $U(1)_e$ global
symmetry in the bulk, and the bulk remains fully gapped for all
excitations, $i.e.$ there is no Goldstone mode in the bulk at
all~\cite{maxashvin}. Also, following the same argument as in
Ref.~\onlinecite{maxashvin}, a $2\pi-$flux of $a_\mu$ at the
boundary will be screened by $D$ in the bulk, which attaches the
flux with $U(1)_e$ charge $1$. Thus the gauge field $a_\mu$
becomes a noncompact gauge theory at the $2d$ boundary, because
its flux now carries a conserved $U(1)_e$ charge, which is
precisely described by the last term of Eq.~\ref{qed2}.


Based on the argument above, the $(2+1)d$ boundary of the system
is described by Eq.~\ref{qed2} because the spinon $\psi_j$ carries
$U(1)_s$ charge, and the gauge flux of $a_\mu$ carries unit
$U(1)_e$ charge. If we break the time-reversal symmetry at the
boundary, $\psi_j$ will acquire a mass term, which will generate a
Chern-Simons term for both $a_\mu$ and $A^s_\mu$ at level $+1$.
Now after integrating out $a_\mu$, the external field $A^{e}_\mu$
will acquire a CS term at level $-1$. The full response theory at
the boundary reads: \beqn \mathcal{L} = \frac{i}{4\pi} A^e \wedge
dA^e - \frac{i}{4\pi} A^s \wedge dA^s. \eeqn This response theory
has already been derived in Ref.~\onlinecite{senthilashvin} for
the boundary of $3d$ bosonic SPT states. This response theory is
consistent with the physics of bosonic TI: it is fully gapped and
has no fractional excitations in the bulk, but if time-reversal
symmetry is broken at the boundary, the boundary will be driven to
a quantum Hall state with Hall conductivity $\sigma_{xy} = \pm 1$.
This also implies that the bulk response theory to $A^e_{\mu}$ and
$A^s_{\mu}$ will acquire a topological term $\theta \vect{E}\cdot
\vect{B}/(4\pi^2)$ term with $\theta = \pm 2\pi$ respectively.

{\it --- 3. Self-duality of the boundary theory }

Now we argue that Eq.~\ref{qed2} has a self-dual structure. Since
in our system $\psi_1$ and $\psi_2$ each has its own U(1) global
symmetry $\psi_j \rightarrow \psi_j e^{i\theta_j}$, and they come
from two independent $n=1$ TIs (let us tentatively ignore the
gauge field $a_\mu$ they couple together), let us form independent
superconductor Cooper pair condensate $\psi^t_j \sigma^y \psi_j
\sim \Delta_j \sim \exp(i\phi_j)$, where $\sigma^y = \gamma^0$ in
Eq.~\ref{qed2}. We can destroy the superconductors by
proliferating the vortices of the superconductors. But here we
would like to consider the quartic vortex of $\phi_1$ and $\phi_2$
individually, namely vortices of $\phi_j$ that $\psi_j$ would view
as a $4\pi$ flux. As was shown in
Ref.~\onlinecite{maxashvin,wangsenthil1,wangsenthil2}, the charge
neutral quartic vortex in a $n=1$ TI is a fermion. This can be
understood by gauging the global U(1) symmetry of $\psi_j$, and
consider the statistics of the $(0, 4\pi)$ monopole. The $(0,
4\pi)$ monopole is naturally a bound state of $(1/2, 2\pi)$ and
$(-1/2, 2\pi)$ dyons, hence it carries angular momentum $1/2$, and
it is a Kramers doublet fermion~\cite{wangpottersenthil}. After
the proliferation of these fermionic vortices, the dual boundary
theory in terms of these fermionic vortices
reads~\cite{maxashvin,wangsenthil1,wangsenthil2}: \beqn
\mathcal{L} = \sum_{j=1}^2 \bar{\chi}_j \gamma_\mu (\partial_\mu -
4 i a^{(j)}_{\mu} )\chi_j + \cdots \label{dual}\eeqn Here $\chi_j$
is the dual Kramers doublet fermion that transform as
$\mathcal{T}: \chi_j \rightarrow i \sigma^y \chi_j$. $a_\mu^{(j)}$
corresponds to the Goldstone mode of $\phi_j$: $\partial_\mu
\phi_j = \frac{1}{2\pi} \epsilon_{\mu\nu\rho}
\partial_\nu a^{(j)}_\rho$.

Now we turn back on the original gauge field $a_\mu$. In the
superconductor phase, the low energy Lagrangian for the two
superconductors that couple to $a_\mu$ is: \beqn \mathcal{L} =
\sum_{j = 1}^2 - t (\partial_\mu \phi_j - 2 a_\mu + (-1)^j 2
A^s_\mu)^2 + \frac{i}{2\pi} \epsilon_{\mu\nu\rho} A^e_\mu
\partial_\nu a_\rho.  \eeqn After going through the standard duality
formalism, we obtain the following Lagrangian: \beqn \mathcal{L} =
\sum_{j = 1}^2 \frac{2i}{2\pi} \epsilon_{\mu\nu\rho} a^{(j)}_\mu
\partial_\nu (a_\rho + (-1)^j A^s_\mu ) + \frac{i}{2\pi}
\epsilon_{\mu\nu\rho} A^e_\mu
\partial_\nu a_\rho.  \eeqn Integrating out $a_\mu$ will generate
the following constraint: \beqn 2 a_\mu^{(1)} + 2 a_\mu^{(2)} +
A^e_\mu = 0, \eeqn or in other words the photon phase of $a_\mu$
will ``Higgs" and gap out the mode $2 a_\mu^{(1)} + 2 a_\mu^{(2)}
+ A^e_\mu$. This constraint can be solved by introducing a new
gauge field $c_\mu$: $4 a_\mu^{(1)} = - c_\mu + A^e_\mu $, $4
a_\mu^{(2)} = c_\mu + A^e_\mu $. Plugging these fields in to
Eq.~\ref{dual}, we obtain the full dual theory of Eq.~\ref{qed2}:
\beqn \mathcal{L} &=& \sum_{j=1}^2 \bar{\chi}_j \gamma_\mu
(\partial_\mu - i (-1)^j c_{\mu} - i A^e_\mu )\chi_j \cr\cr &+&
\frac{i}{2\pi} \epsilon_{\mu\nu\rho} c_\mu \partial_\nu A^s_\rho.
\label{dual2}\eeqn We can see that $\chi_j$ are two Dirac fermions
that each carries $U(1)_e$ charge $+1$. They correspond to the
quartic vortex of the superconductor of $\psi_j$. Physically this
is easy to understand: $\chi_j$ is a quartic vortex of $\phi_j$,
and a quartic vortex of $\phi_j$ carries gauge flux $2\pi$ of
$a_\mu$, which due to the last term of Eq.~\ref{qed2} should also
carry global $U(1)_e$ charge 1. Notice that here we create quartic
vortex of $\phi_1$ and $\phi_2$ individually, namely a vortex of
$\phi_1$ alone without a vortex of $\phi_2$ will carry $a_\mu$
gauge flux $2\pi$.

According to Eq.~\ref{dual2}, the flux of $c_\mu$ carries unit
$U(1)_s$ charge. Again this can be physically understood as
following: a flux of $c_\mu$ is the difference between the flux
number of $a^{(1)}_\mu$ and $a^{(2)}_\mu$, and based on the
standard boson-vortex duality, the flux of $c_\mu$ also
corresponds to the density difference between $\psi_1$ and
$\psi_2$, which is a quantity that does not carry $U(1)_e$ charge,
but carries $U(1)_s$ charge. Now after a particle-hole
transformation $\chi_2 \rightarrow \chi_2^\dagger$, the dual
boundary theory Eq.~\ref{dual2} takes exactly the same form as
Eq.~\ref{qed2}, with an exchanged role between $A^e_\mu$ and
$A^s_\mu$: \beqn \mathcal{L}_{dual} &=& \sum_{j = 1}^2
\bar{\chi}_j \gamma_\mu (\partial_\mu - i c_{\mu}) \chi_j \cr\cr
&-& i A^e_\mu \bar{\chi}_i \gamma_\mu \tau^z_{ij}\chi_j +
\frac{i}{2\pi} \epsilon_{\mu\nu\rho} c_\mu
\partial_\nu A^s_\rho. \label{dual3}\eeqn

Because the dual fermion $\chi_j$ transforms as $\mathcal{T}:
\chi_j \rightarrow i\sigma^y \chi_j$ under time-reversal, if we
ignore the gauge field $c_\mu$ in Eq.~\ref{dual2}, the symmetry
for $\chi_j$ is $U(1)_e \rtimes Z_2^T$, which is the symmetry of
the ordinary $3d$ TI~\cite{moorebalents2007,fukane,roy2007}, and
as is well-known, it has an $\mathbb{Z}_2$ classification. This
implies that without $c_\mu$, there is a mass term that is allowed
by all the symmetries: $m \chi^\dagger_{i} \sigma^y \otimes
\tau^y_{ij} \psi_j $. However, because $\chi_1$ and $\chi_2$ carry
opposite gauge charge under $c_\mu$ in Eq.~\ref{dual2}, this mass
term is forbidden by the gauge symmetry. Thus this dual boundary
theory Eq.~\ref{dual2} cannot be trivially gapped out without
breaking symmetry or gauge symmetry.

If we explicitly break either the $U(1)_e$ or $U(1)_s$ symmetry,
then the $3d$ bulk can be called a bosonic TI. But in this case
$a_\mu$ or $c_\mu$ will become a compact gauge field, because
their fluxes will no longer carry conserved quantities, hence the
instanton monopole process is allowed in the $(2+1)d$ space-time.
And inspired by Ref.~\onlinecite{mirror}, we conjecture that the
duality we propose here may have an analogue in supersymmetric
field theories.

{\it --- 4. Relation to other possible boundary states}

There are many possible boundary states of this system, for
example states with different spontaneous symmetry breaking. We
are most interested in boundary states that do not break any
symmetry. Ref.~\onlinecite{senthilfisher} gave us another way of
looking at QED$_3$ with $N_f = 2$: Eq.~\ref{qed2} and
Eq.~\ref{dual3} can both be mapped to a O(4) nonlinear sigma model
with a topological $\Theta-$term at $\Theta = \pi$. Here we
reproduce the discussion in Ref.~\onlinecite{senthilfisher}. First
we couple Eq.~\ref{qed2} to a three component dynamical unit
vector field $\vect{N}(x, \tau)$: \beqn \mathcal{L} &=& \sum_{j =
1}^2 \bar{\psi}_j \gamma_\mu (\partial_\mu - i a_\mu) \psi_j + m
\bar{\psi} \vect{\tau} \psi \cdot \vect{N}, \eeqn introducing this
slow moving vector $\vect{N}$ is equivalent to turning on certain
four fermion interaction for $\psi_j$, and $\vect{N}$ could be
introduced through Hubbard-Stratonovich transformation.

Now following the standard $1/m$ expansion of
Ref.~\onlinecite{abanov2000}, we obtain the following action after
integrating out the fermion $\psi_j$: \beqn \mathcal{L}_{eff} =
\frac{1}{g} (\partial_\mu \vect{N})^2 + i \pi
\mathrm{Hopf}[\vect{N}] + i a_\mu J_\mu^T + \frac{1}{e^2}
f_{\mu\nu}^2, \label{hopf} \eeqn where $1/g \sim m$. $J_0^T =
\frac{1}{4\pi} \epsilon_{abc} N^a
\partial_x N^b \partial_y N^c$ is the Skyrmion density of $\vect{N}$,
thus $J_\mu^T$ is the Skyrmion current. The second term of
Eq.~\ref{hopf} is the Hopf term of $\vect{N}$ which comes from the
fact that $\pi_3[S^2] = \mathbb{Z}$.

Now if we introduce the CP$^1$ field $z_\alpha = (z_1, z_2)^t =
(n_1 + i n_2, n_3 + in_4)^t$, the Hopf term becomes precisely the
$\Theta-$term for the O(4) vector $\vect{n}$ with $\Theta = \pi$:
\beqn i \pi \mathrm{Hopf}[\vect{N}] = \frac{i \pi}{2\pi^2}
\epsilon_{abcd} n^a
\partial_x n^b \partial_y n^c
\partial_\tau n^d. \label{o4}\eeqn Here $\Theta = \pi$ is protected by
time-reversal symmetry. This is because $\vect{N}$ is odd under
time-reversal, and hence $z_\alpha$ is a Kramers doublet boson.
Simple algebra shows that Eq.~\ref{o4} changes sign under
time-reversal. In the CP$^1$ formalism, the Skyrmion current
$J^T_\mu = \frac{1}{2\pi} \epsilon_{\mu\nu\rho}\partial_\nu
\alpha_\rho$, where $\alpha_\mu$ is the gauge field that the
CP$^1$ field $z_\alpha$ couples to. Due to the coupling $a_\mu
J^T_\mu = \frac{i}{2\pi} \epsilon_{\mu\nu\rho}a_\mu
\partial_\nu \alpha_\mu$, after integrating out
$a_\mu$, $\alpha_\mu$ is Higgsed and gapped, and $z_\alpha$
becomes a complex boson that does not couple to any gauge field,
and its transformation $z_\alpha \rightarrow e^{i\theta} z_\alpha$
becomes the physical $U(1)_e$ symmetry, due to the mutual
Chern-Simons coupling between $a_\mu$ and $\alpha_\mu$. Thus
$z_\alpha$ now carries both the $U(1)_e$ and $U(1)_s$ quantum
numbers, and if we start with the dual theory Eq.~\ref{dual3}, the
same O(4) NLSM with $\Theta-$term in Eq.~\ref{o4} can be derived.

The phase diagram of the O(4) NLSM with a $\Theta-$term was
discussed in Ref.~\onlinecite{xuludwig}, and it was proposed that
in the large $g$ (small $m$) disordered phase, $\Theta = \pi$ is
the quantum critical point (quantum phase transition) between
stable fixed points $\Theta = 0$ and $\Theta = 2\pi$, which is
consistent with the conjecture made in
Ref.~\onlinecite{senthilfisher} that the quantum disordered phase
of the O(4) NLSM with $\Theta=\pi$ could be a gapless paramagnet
(a 2+1d CFT). Recently this conjecture was confirmed numerically
in Ref.~\onlinecite{kevinQSH,mengQSH2}, and the sign-problem-free
simulation in both Ref.~\onlinecite{kevinQSH,mengQSH2} strongly
suggest that the quantum disordered phase of the O(4) NLSM with
$\Theta = \pi$ is indeed a strongly coupled CFT, sandwiched
between two fully gapped quantum disordered phases controlled by
fixed points $\Theta = 0$ and $2\pi$ (Fig.4 in
Ref.~\onlinecite{mengQSH2}, and discussion therein).

The $1/m$ expansion above is certainly valid for large $m$ (small
$g$), which corresponds to the ordered phase of the O(4) CP$^1$
field $\vect{n}$ and three component vector $\vect{N}$. The usual
expectation of QED$_3$ with $N_f = 2$ is that it leads to
spontaneous chiral symmetry breaking at low
energy~\cite{qed1,qed2,qed3,qed4}, which precisely corresponds to
the order of vector $\vect{N}$. However, if a proper four fermion
interaction term is turned on in Eq.~\ref{qed2} and
Eq.~\ref{dual3} that prevents the chiral symmetry breaking, it may
remain a CFT that corresponds to the disordered phase of O(4) NLSM
with $\Theta = \pi$.



Other possible boundary states can be constructed through the O(4)
NLSM with $\Theta = \pi$, as was discussed in
Ref.~\onlinecite{senthilashvin,xuclass}. For example, let us break
the $U(1)_s$ symmetry, and keep the following time-reversal
symmetry $\mathcal{T}: z_\alpha \rightarrow (\tau^x)_{\alpha\beta}
z_\beta =  \left( \tau^y \exp( - i \tau^z \pi/2 )
\right)_{\alpha\beta} z_\beta $, then one can see that this O(4)
NLSM model with $\Theta = \pi$ precisely correspond to the
boundary of the bosonic TI with $U(1)_e \rtimes Z_2^T$ symmetry.
And following the discussion in Ref.~\onlinecite{senthilashvin},
this $\Theta-$term can drive the boundary into the so called
$eCmC$ $Z_2$ topological order, namely its $e$ and $m$ anyons with
mutual semion statistics both carry half $U(1)_e$ charge.

The authors are supported by the David and Lucile Packard
Foundation and NSF Grant No. DMR-1151208. The authors are grateful
to Chong Wang, T. Senthil for very helpful discussions. We also
acknowledge a related unpublished work we learned through private
communication~\cite{private}.

\bibliography{BTI}

\end{document}